\documentclass[aps]{revtex4}
\ifx\pdfoutput\undefined
\usepackage[dvips]{graphicx}
\else
\usepackage[pdftex]{graphicx}
\fi
\usepackage[applemac]{inputenc}
\usepackage{amssymb}

\def\bx{{\bf x}}
\def\br{{\bf r}}
\def\brp{{\bf {r^\prime}}}
\def\bxp{{\bf x^\prime}}

\def\ap{{\alpha^{\prime}}}
\def\bp{{\beta^{\prime}}}

\begin{document}

\title{ Superoperator Many-body Theory of Molecular Currents:
Non-equilibrium Green Functions in Real Time }
\author{ U. Harbola$^\dagger$ and S. Mukamel$\bf^*$\\
Department of Chemistry, University of California, Irvine, California 92697-2025, United States }

\begin{abstract}
The electric conductance of a molecular junction is calculated by recasting the Keldysh
formalism in Liouville space. Dyson equations for
nonequilibrium many body Green's functions (NEGF) are derived directly in real (physical) time.
The various NEGFs appear naturally in the theory as time ordered products of superoperators, while
the Keldysh forward/backward time loop is avoided.

\vspace{1cm}
\noindent PACS No: 73.23.-b, 68.43.pq, 05.60.Gg
\end{abstract}

{\it Submitted to Phys. Rev. B}

\maketitle

\section*{I. Introduction}
Recent advances in the fabrication and measurments of nanoscale devices have lead to a considerable
interest in nonequilibrium current carrying states of single molecules. The tunneling of electrons between two metals separated by a thin oxide layer was first observed experimentally by Giaever \cite{giaever} and later by others  \cite{nicol}.  Vibrational resonances can
be observed for molecules absorbed at the metal-oxide interface by analyzing the tunneling current
as a function of the applied bias\cite{jaklevik,reed}. More recent development of Scanning Tunneling Microscopy (STM) lead to a direct, real space determination of surface structures.
A metal tip is brought near the surface so that tunneling resistence is measurable. A contour map of the surface is obtained by recording the tunneling resistance as the tip scans the surface. The tunneling electrons interact and may exchange energy with the nuclear degrees of freedom of the absorbed molecule.
This opens up inelastic channels for electron transmission from tip to the surface  leading to the
ineastic electron tunneling (IET).  IET may play an important role in manipulating molecules
with STM\cite{walkup,gao}.
Recently, the IET was combined with STM for the chemical analysis of a single absorbed molecule with
atomic spatial resolution \cite{ho,stipe}. Recent advances in the theory of STM are reviewed in Ref. \cite{hofer}.

Electron tunneling was first analyzed by Bardeen and Cohen $et$ $al$ \cite{bardeen,mhcohen}
using the perturbative transfer Hamiltonian (TH) approach  and more recently by many other
authors \cite{tersoff,lang,perssonprl1987}.
Although the TH gives, in most cases, a good description of the observed effects, it lacks a firm first principles theoretical basis and does not account properly for many-body effects \cite{brinkman}.
An improved form of TH \cite{brinkman1} that involved energy dependent transfer matrix elements, was used to
incorporate many body effects. However this model does not describe the elecron-phonon interaction
properly\cite{caroli1}.

A many-body non-equilibrium Green functions (NEGF) formulation of electron tunneling
was proposed by Caroli $et$ $al$ \cite{caroli}.
The NEGF theory was originated by Schwinger \cite{schwinger1961}
and Kadanoff and Baym \cite{kad-baym-1962}, and developed further by Keldysh \cite{keldysh}
and Craig \cite{craig1968}. This formalism involves the calculation of four basic Green functions, time ordered ($G^T$), anti-time ordered ($G^{\tilde T}$), lesser ($G^<$) and greater ($G^>$).
Additional retarded ($G^r$) and advanced ($G^a$) Green's functions are defined as
specific combinations of these basic functions.
At equilibrium suffice it to know only the retarded or advanced Green functions; all
other Green functions simply follow from the fluctuation-dissipation theorem that connects the
"lesser" and "greater"  with the retarded Green function through the equilibrium fermi distribution function ($f_0(E)$) \cite{haug-jauho}. However, for nonequilibrium measurements, where the distribution function is not known a priory, one needs to solve for the various NEGFs self-consistently.

Electronic transport in molecular wires and STM currents of single molecules have received
considerable attention \cite{a-b-nitzan,datta,ventra,tian,baer,ness}.
Electron transport through a single molecule \cite{science1,science2,park} or a chain of several atoms \cite{agrait} were studied. From a theoretical point of view, this is very similar to the electron tunneling in semiconductor junctions and various theories developed for STM \cite{datta-book,anitzan3} can directly be applied to molecular wires.
The NEGF technique developed for tunneling currents has been used to analyze the electron conduction
through a single molecule attached to electrodes
\cite{tikhodeev2001,tikhodeev,dicarlo,a-b-nitzan,799,mii2003,datta-book}.
The method has also been combined with density functional theory for the modeling of
transport in molecular devices \cite{datta2002,oleg}.

In this paper we develop a nonequilibrium superoperator Green function theory \cite{keldysh,haug-jauho,hao}
(NESGF) of molecular currents \cite{Mukamel}.
A notable advantage of working with superoperators in the higher dimensional
Liouville space  \cite{adam-mukamel,oleg}
is that we need only consider time ordered quantities in real (physical) time; all
NEGFs show up naturally without introducing artificial time variables. Observables can be expressed
in terms of various {\it Liouville space pathways}(LSP) \cite{Mukamel}.
The ordinary (causal) response function which represents the density response to an external field
is one particular combination of these LSPs. Other combinations represent the spontaneous density
 fluctuations and the response of these fluctuations to the external field \cite{adam-mukamel,wang}.
A simple time ordering operation of superoperators in real time is all it takes to derive the non-equiliubrium theory avoiding the Keldysh loop or Matsubara imaginary time.
The NESGF theory provides new physical insights into the mechanism of the current. It can also
be more naturally used to interpret time domain experiments involving external pulses.

In Sec. II, we give a brief introducion to the superoperator formalism and recast the NEGF theory
in terms of the superoperator Green functions. Starting from the microscopic definitions
for various non-equilibrium superoperator Green functions (NESGF), we construct dynamical
equations of motion and obtain the Dyson matrix equation of Keldysh which couples the various NESGFs. In Sec. III, we apply the NESGF theory to the cunduction of a molecular junction.
In Sec. IV  we end with a discussion.

\section*{II. Dyson Equations for Superoperator Green Functions}
We consider a system of externally driven  electrons and phonons
described by the Hamiltonian \cite{keldysh,perssonprl1987,tikhodeev2001,a-b-nitzan},
\begin{equation}
\label{hamilton}
H= H_0 + H_{ep} + H_{ex}
\end{equation}
where $H_0$ represents the non-interacting electrons and phonons,
\begin{equation}
\label{h0}
H_0= \int d\br \psi^\dagger(\br) h_0(\br) \psi(\br)
 + \int d\br \phi^\dagger(\br) \Omega_0(\br) \phi(\br).
\end{equation}
$h_0(\br)=(-\hbar^2/2m) {\bf \nabla}^2$ is the kinetic energy and $m$ is the electron mass.
$\psi$ ($\psi^\dagger$) represent the anihilation (creation)
operators which satisfy the fermi anticommutation relations,
\begin{eqnarray}
\label{usual-relations}
\psi(\br)\psi^\dagger(\brp) + \psi^\dagger(\brp) \psi(\br) &=&
\delta(\br-\brp)\nonumber\\
\psi^\dagger(\br)\psi^\dagger(\brp) + \psi^\dagger(\brp)
\psi^\dagger(\br) &=& 0 \nonumber\\
\psi(\br)\psi(\brp) + \psi(\brp) \psi(\br) &=& 0
\end{eqnarray}
and $\phi$ ($\phi^\dagger$) are boson operators with the commutaion relations,
\begin{eqnarray}
\label{usual-relations1}
\phi(\br)\phi^\dagger(\brp) - \phi^\dagger(\brp) \phi(\br) &=&
\delta(\br-\brp)\nonumber\\
\phi^\dagger(\br)\phi^\dagger(\brp) - \phi^\dagger(\brp)
\phi^\dagger(\br) &=& 0 \nonumber\\
\phi(\br)\phi(\brp) - \phi(\brp) \phi(\br) &=& 0
\end{eqnarray}
The second term in Eq. (\ref{hamilton}) denotes the electron - phonon
interaction,
\begin{equation}
H_{ep} = \int d\br \lambda(\br) [\phi^\dagger(\br)+\phi(\br)] \psi^\dagger(\br) \psi(\br).
\end{equation}
where $\lambda(\br)$ is the coupling strength.
 Finally, $H_{ex}$ represents the coupling to a time dependent external potantial $\xi(\br,t)$,
\begin{equation}
\label{ex-hamil}
H_{ex} = \int d\br \xi(\br,t) \psi^\dagger(\br) \psi(\br)
\end{equation}

We next briefly survey some properties of Liouville space superoperators
that will be useful in the following derivations \cite{shaul-pre68}.
The elements of the Hilbert space $N \times N$ density matrix,
$\rho(t)$, are arranged as a Liouville space vector (bra or ket) of length $N^2$.
Operators of  $N^2\times N^2$ dimension  in this space are denoted as
superoperators.
With any Hilbert space operator $A$, we associate two superoperators
$A_L$ (left) and $A_R$ (right) defined through their action on another operator $X$
as,
\begin{equation}
\label{def-1}
A_L X \equiv A X ~~~~ \mbox{and} ~~~~ A_R X \equiv X A.
\end{equation}
We further define  symmetric and antisymmetric combinations of these superoperators,
\begin{equation}
\label{def-2}
A_+ = \frac{1}{2} ( A_L + A_R ) ~~~~~ \mbox{and} ~~~~~  A_- =  ( A_L - A_R ).
\end{equation}
The commutator and anticommutator operations in Hilbert space can thus be implemented
with a single multiplication by a "-" and "+" superoperators, respectively.
We further introduce the Liouville space time ordering operator ${\cal T}$.
This is a key ingredient for extending NEGF to superoperators: when
applied to a product of superoperators it reorders them so that time increases from right to left.
We define $\langle A(t) \rangle $ $\equiv$ Tr$\{ A(t) \rho_{eq}\}$, where $\rho_{eq}=\rho(t=0)$ represents the equilibrium density matrix of the interacting system.
It is straightforward to see that for any two operators $A$ and $B$ we have,
\begin{equation}
\label{ret-adv}
\langle {\cal T} A_+(t) B_-(t^\prime) \rangle =  0  ~~~~ t^\prime > t
\end{equation}
$\langle {\cal T} A_+(t) B_-(t^\prime) \rangle$ is thus always a retarded function.
This follows from the definitions (\ref{def-2}). Since a "-" superoperator corresponds to a commutator in Hilbert space, this implies that for $t<t^\prime$, $\langle A_+(t) B_-(t^\prime) \rangle$  becomes a trace
 of a commutator and must vanish, $i.$$e.$,
\begin{eqnarray}
\langle {\cal T} A_+(t) B_-(t^\prime) \rangle  &=& \mbox{Tr}\{B_-(t^\prime) A_+(t) \rho_{eq}\} ~~~~ t<t^\prime \nonumber\\
&=& \frac{1}{2} Tr\{[B(t^\prime), A(t)\rho_{eq}+\rho_{eq}A(t)] \} = 0 \nonumber
\end{eqnarray}
 Similarly, it follows that the trace of two "minus" operators always vanishes,
\begin{equation}
\label{ret-adv1}
\langle {\cal T} A_-(t) B_-(t^\prime) \rangle =  0 ~~~~ \mbox{for all
t and $t^\prime$}.
\end{equation}
We shall make use of Eqs. (\ref{ret-adv}) and (\ref{ret-adv1})
for discussing the retarded and advanced Green functions in Appendix D. Superoperator algebra
was surveyed in Ref. \cite{shaul-pre68}.

In Liouville space the density matrix, $\rho(t)$ is a vector whose time dependence
is given by,
\begin{equation}
\label{rho}
\rho(t)= {\cal G}(t,t_0)  \rho(t_0)
\end{equation}
with the Green function,
\begin{equation}
\label{time-1}
{\cal G}(t,t_0) = {\cal
T}\mbox{exp}\left\{-\frac{i}{\hbar}\int_{t_o}^t{\cal H}_- (\tau) d\tau\right\},
\end{equation}
and ${\cal H}_-$ is the superoperator corresponding to the
Hamiltonian (Eq. \ref{hamilton}).
Note that unlike Hilbert space, where time dependence of the ket
and the bra is governed by forward and backward time-evolution operators respectively, in
Liouville space one keeps track of both bra and ket simultaneously and the density matrix
needs only to be propagated forward in time ( Eq. (\ref{rho})).

To introduce the interaction picture in Liouville space we
partition ${\cal H}_- = {{\cal H}_0}_- + {{\cal H}^\prime}_-$  corresponding
to the non-interacting and interaction Hamiltonians. With this partitinaing, Eq. (\ref{time-1}) can
 be written as,
\begin{equation}
\label{inter-G}
{\cal G}(t,t_0) = {\cal G}_0(t,t_0) {\cal G}_I(t,t_0)
\end{equation}
where ${\cal G}_0$ represents the time evolution with respect to $H_0$,
\begin{equation}
{\cal G}_0(t,t_0)= \theta(t-t_0)\mbox{exp}\left\{-\frac{i}{\hbar}{{\cal H}_0}_-(t-t_0) \right\}.
\end{equation}
${\cal G}_I(t,t_0)$ is the time evolution operator in the
interaction picture,
\begin{equation}
\label{time-int}
{\cal G}_I(t,t_0)= {\cal
T}\mbox{exp}\left\{-\frac{i}{\hbar}\int_{t_o}^t {\tilde{\cal H}}^\prime_-(\tau) d\tau\right\}
\end{equation}
and ${\tilde{\cal H}}^\prime_-$ is the interaction picture representation of
${\cal H}^\prime_-$. We shall denote superoperators in the interaction picture by a
$(~\tilde{}~)$,
\begin{equation}
\label{time-L}
\tilde{A}_\alpha(t) \equiv {\cal G}^\dagger_0(t,t_0) {A}_\alpha(t_0) {\cal G}_0(t,t_0)
\end{equation}
where $\alpha$= +,- or $L$,$R$. Superoperators in the Heigenberg picture will be represented
by a caret,
\begin{equation}
\label{heisen}
\hat{A}_\alpha(t) \equiv {\cal G}^\dagger(t,t_0) {A}_\alpha(t_0) {\cal G}(t,t_0)
\end{equation}

By adiabatic switching of the interaction ${{\cal H}^\prime}_-$ starting at $t_0=-\infty$ we have,
\begin{equation}
\label{rhot}
\rho(t)= \rho_0 -\frac{i}{\hbar} \int^t_{-\infty} d\tau {\cal
G}_0(t,\tau){{\cal H}^\prime}_-(\tau) \rho(\tau)
\end{equation}
where $\rho_0=\rho(-\infty)$ is the equilibrium density matrix of the
non-interacting system,
\begin{equation}
\rho_0 = \frac{\mbox{exp}(-\beta H_0)}{\mbox{Tr}\{{\mbox{exp}(-\beta H_0)}\}}
\end{equation}
An iterative solution of Eq. (\ref{rhot}) yields,
\begin{equation}
\label{rhot-1}
\rho(t)=  {\cal G}_I(t,-\infty) \rho_0
\end{equation}
which can also be obtained by applying the time evolution operator
(\ref{time-int}) to $\rho(t_0)$ and setting $t_0=-\infty$.
Using Eq. (\ref{rhot-1}), the equilibrium density matrix of the interacting system can be
generated from the non-interacting one by switching on the interactions adiabatically starting
at $t=-\infty$.
The external potential is constant in time for $t < 0$ and is assumed to be time dependent
only for $t>0$. We then get,
\begin{equation}
\label{rhot-2}
\rho_{eq}=  {\cal G}_I(0,-\infty) \rho_0,
\end{equation}
This {\em adiabatic connection formula } has been shown \cite{shaul-pre68} to be very
useful for calculating expectaion values using the interaction picture. In the corresponding Gellman-law expression in Hilbert space \cite{negele} there is an extra denominator that takes
care of the phase of the wavefunction.
This is not necessery in Liouville space since the density  matrix does not acquire such a phase.

In the Heisenberg picture, the expectaion value of an operator $\hat{A}_{\alpha}(t)$ is given by,
\begin{equation}
\label{heisen-form}
\langle \hat{A}_\alpha(t) \rangle \equiv \mbox{Tr}\left\{ \hat{A}_\alpha(t)
\rho_{eq}\right\},
\end{equation}
where $\rho_{eq}=\rho(t=0)$.
Using Eqs. (\ref{time-L}) (\ref{heisen}) and (\ref{rhot-1}), this can be recast in the
interaction picture as,
\begin{equation}
\label{perturb}
\langle \hat{A}_\alpha(t) \rangle = \mbox{Tr}\left\{ \tilde{A}_\alpha(t)
{\cal G}_I(t,-\infty) \rho_0 \right\}\equiv \langle \tilde{A}_\alpha(t){\cal G}_I(t,-\infty)\rangle_0.
\end{equation}
Equation (\ref{perturb}) is a good starting point for developing a perturbation theory around the
 non-interacting system. Through Eqs. (\ref{heisen-form}) and (\ref{perturb}) we also define the
 expectation values $\langle ... \rangle$ and $\langle...\rangle_0$.
 While the former represents the trace
 with respect to the interacting density matrix, the latter is defined with respect to the non-interacting
 density matrix. This will be used in the following.

Corresponding to the Hilbert space electron and phonon
operators, $\hat{\psi}$, $\hat{\psi}^\dagger$, $\hat{\phi}$ and $\hat{\phi}^\dagger$
we define "left" ($\alpha$=$L$)and "right" ($\alpha$=$R$) superoperators, $\hat{\psi}_\alpha$,
$\hat{\psi}^\dagger_\alpha$, $\hat{\phi}_\alpha$ and $\hat{\phi}^\dagger_\alpha$.
The dynamics of a superoperator, $\hat{\psi}_\alpha$, is described by the
generalized Liouville equation,
\begin{equation}
\label{dyn-eq}
-i\hbar \frac{\partial \hat{\psi}_\alpha(t)}{\partial t} = [{\cal H}_-(t),
\hat{\psi}_\alpha(t)] = {\cal H}_-(t) \hat{\psi}_\alpha(t) - \hat{\psi}_\alpha(t) {\cal H}_-(t),
\end{equation}
where  ${\cal H}_-$ is the superoperator corresponding to the Hamiltonian
given in Eq. (\ref{hamilton}).
A similar equation can be written down for the phonon superoperators.
In order to evaluate the commutator appearing in the RHS of Eq.
(\ref{dyn-eq}), we need the commutation relations of superoperators \cite{thomas-shaul}.
The "left" and the "right" operators always commute. Thus for $\alpha\neq \beta$ we have,
\begin{eqnarray}
\label{commutation1}
[\psi_\alpha(\br), \psi_\beta(\brp)] &=&
[\psi^\dagger_\alpha(\br), \psi^\dagger_\beta(\brp)]
= [\psi^\dagger_\alpha(\br), \psi_\beta(\brp)] = 0  \nonumber\\
\left[\phi_\alpha(\br), \phi_\beta(\brp)\right] &=&
[\phi^\dagger_\alpha(\br), \phi^\dagger_\beta(\brp)]
= [\phi^\dagger_\alpha(\br), \phi_\beta(\brp)] = 0
\end{eqnarray}
For fermi superoperators we have,
\begin{eqnarray}
\label{commutation}
\psi_\alpha(\br) \psi_\alpha(\brp) + \psi_\alpha(\brp)
\psi_\alpha(\br) &=& 0 \nonumber\\
\psi^\dagger_\alpha(\br) \psi^\dagger_\alpha(\brp) +
\psi^\dagger_\alpha(\brp) \psi^\dagger_\alpha(\br) &=& 0
\nonumber\\
\psi_\alpha(\br) \psi^\dagger_\alpha(\brp) + \psi_\alpha(\brp)
\psi^\dagger_\alpha(\br)
 &=& \delta(\br-\brp).
 \end{eqnarray}
Similarly for the boson operators,
\begin{eqnarray}
\label{commutation0}
\phi^\dagger_\alpha(\br) \phi^\dagger_\alpha(\brp) -
\phi^\dagger_\alpha(\brp) \phi^\dagger_\alpha(\br) &=& 0
\nonumber\\
\phi_\alpha(\br) \phi_\alpha(\brp) - \phi_\alpha(\brp)
\phi_\alpha(\br) &=& 0 \nonumber\\
\phi_\alpha(\br) \phi^\dagger_\alpha(\brp) -
\phi^\dagger_\alpha(\brp)
\phi_\alpha(\br) &=& \kappa_\alpha\delta(\br-\brp).
\end{eqnarray}
Here $\kappa_\alpha$= -1 for $\alpha= R$ and unity for $\alpha$ = L.

 Using the commutaion relations (\ref{commutation1}), (\ref{commutation}) and the identity,
\begin{eqnarray}
\label{identity}
( X Y... Z)_\alpha = X_\alpha Y_\alpha... Z_\alpha, ~~~~\alpha=L,R
\end{eqnarray}
we can recast ${\cal H}_-$ in terms of the elementary field superoperators,
\begin{eqnarray}
\label{super-hamil}
 {\cal H}_-= {\cal H}_{0-} + {\cal H}^{ep}_{-} + {\cal H}^{ex}_{-}
\end{eqnarray}
with
\begin{eqnarray}
\label{super-hamil2}
{\cal H}_{0-} &=& \sum_{\alpha = L,R} \kappa_\alpha \int d\br \left(
\psi^\dagger_\alpha(\br) h_0(\br) \psi_\alpha(\br) +
\phi^\dagger_\alpha(\br) \Omega_0(\br) \phi_\alpha(\br) \right) \nonumber\\
{\cal H}^{e-p}_- &=&  \sum_{\alpha=L,R}\kappa_\alpha  \int d\br
\lambda(\br)\Phi_\alpha(\br) \psi^\dagger_\alpha(\br) \psi_\alpha(\br) \nonumber\\
{\cal H}^{ex}_- &=& \sum_{\alpha= L,R} \kappa_\alpha \int d\br
\psi^\dagger_\alpha(\br) \xi(\br,t)\psi_\alpha(\br)
\end{eqnarray}

We next define  electron and phonon superoperator Greens
functions,
\begin{eqnarray}
\label{green}
G_{\alpha\beta}(\br t,\brp t^\prime) &=& -\frac{i}{\hbar}\langle {\cal T}
\hat{\psi}_\alpha(\br,t) \hat{\psi}^\dagger_\beta(\brp,t^\prime) \rangle
\nonumber\\
D_{\alpha\beta}(\br t,\brp t^\prime) &=& -\frac{i}{\hbar}\langle {\cal T}
\hat{\Phi}(\br,t) \hat{\Phi}^\dagger(\brp,t^\prime) \rangle
\end{eqnarray}
As shown in Ref. \cite{oleg} (see Appendix A),
$G_{LL}$, $G_{RR}$, $G_{LR}$ and $G_{RL}$ respectively coincide with
the standard Hilbert space time ordered
$G^T$, antitime ordered $G^{\tilde T}$, lesser $G^{<}$ and greater
$G^{>}$ Green functions defined on a closed time loop.

Using the commutaion relations (\ref{commutation}),
the Heisenberg equations of motion for superoperators $\hat{\psi}_\alpha(t)$ and
$\hat{\Phi}_\alpha(t)$, $\alpha= L,R$, read
\begin{eqnarray}
\label{eq-motion}
i\hbar\kappa_\alpha\frac{\partial \hat{\psi}_\alpha(\br,t)}{\partial t}&=&
h_(\br,t) \hat{\psi}_\alpha(\br,t) + \lambda(\br) \hat{\Phi}_\alpha(\br,t)
\hat{\psi}_\alpha(\br,t) \nonumber\\
-i\hbar\kappa_\alpha\frac{\partial \hat{\Phi}_\alpha(\br,t)}{\partial t} &=&
\Omega_0(\br) \hat{\Phi}_\alpha(\br,t)
 + \lambda(\br) \hat{\psi}^\dagger_\alpha(\br,t) \hat{\psi}_\alpha(\br,t),
\end{eqnarray}
where $h(\br,t)= h_0(\br)+\xi(\br,t)$.
By taking the time derivaive of the electron Green function in Eq.
(\ref{green}) and using
Eqs. (\ref{eq-motion}), we obtain the equation of motion for
$G_{\alpha\beta}$,
\begin{eqnarray}
\label{dyn-g}
\left(i\hbar \frac{\partial}{\partial t} - \kappa_\alpha h(\br,t)\right)
G_{\alpha\beta}(\br t,\brp, t^\prime) &=& \delta_{\alpha\beta}
\delta({\bf x} - {\bf x}^\prime) - \frac{i}{\hbar}\kappa_\alpha \lambda(\br) \langle {\cal T} \hat{\Phi}_\alpha(\br,t) \hat{\psi}_\alpha(\br,t) \hat{\psi}^\dagger_\beta (\brp,t^\prime) \rangle,
\end{eqnarray}
and similarly for the phonon Green function,
\begin{equation}
\label{dyn-d}
\left( i\hbar \frac{\partial}{\partial t} + \kappa_\alpha
\Omega_{0}(\br)\right)
D_{\alpha\beta}(\br t,\brp, t^\prime) = \delta_{\alpha\beta}
\delta({\bf x} - {\bf x}^\prime) +
\frac{i}{\hbar} \lambda(\br) \kappa_\alpha \langle {\cal T}
\hat{\psi}^\dagger_\alpha(\br,t) \hat{\psi}_\alpha(\br,t)
\hat{\Phi}^\dagger_\beta (\brp,t^\prime) \rangle
\end{equation}
We shall denote the space and time coordinates collectively by $\bx=\br,t$; thus
 in Eqs. (\ref{dyn-g}) and (\ref{dyn-d}) $\delta({\bf x} - {\bf
x}^\prime)\equiv \delta({\bf r} - \brp) \delta(t - t^\prime)$.

Following Keldysh, we shall rearrange the superoperator Green functions in  a
$2\times 2$ matrix $\bar{G}$,
\begin{equation}
\bar{G}(\bx,\bxp) = \left(
\begin{array}{cc}
G_{LL}(\bx,\bxp) & G_{LR}(\bx,\bxp)\\
G_{RL}(\bx,\bxp) & G_{RR}(\bx,\bxp)
\end{array}
\right)
\end{equation}
and similarly the phonon Green function matrix $\bar{D}$ with elements $D_{\alpha\beta}$.
The corresponding Green functions of the noninteracting system described by the Hamiltonian (\ref{h0}) are denoted by $\bar{G}^0$ and $\bar{D}^0$, respectively. These are given by,
\begin{eqnarray}
\label{zero-green}
G^0_{\alpha\beta}(\br t,\brp t^\prime) &=& \left(i\hbar
\frac{\partial}{\partial t} -
\kappa_\alpha h(\br,t)\right)^{-1} \delta_{\alpha\beta}
\delta(\bx-\bxp) \nonumber\\
D^0_{\alpha\beta}(\br t,\brp t^\prime) &=&
\left( i\hbar \frac{\partial}{\partial t} + \kappa_\alpha
\Omega_{0}(\br)\right)^{-1}
\delta_{\alpha\beta} \delta(\bx-\bxp).
\end{eqnarray}
Using our matrix notation, we can recast Eqs. (\ref{dyn-g}), (\ref{dyn-d}) in the form of Dyson equations,
\begin{eqnarray}
\label{dyson}
\bar{G} &=& \bar{G^0} + \bar{G^0}\bar{\Sigma}\bar{G} \nonumber\\
\bar{D} &=& \bar{D^0} + \bar{D^0}\bar{\Pi}\bar{D}.
\end{eqnarray}
The effect of all interactions is now included in the electron
($\bar{\Sigma}$) and phonon ($\bar{\Pi}$)self energies. Exact expressions for the self-energies are
obtained by comparing Eqs. (\ref{dyn-d}) and (\ref{dyn-g}) with Eq. (\ref{dyson}),
\begin{eqnarray}
\label{self-energy}
\Sigma_{\alpha\beta}(\br t,\brp t^\prime) &=& -
\frac{i}{\hbar}\kappa_\alpha \lambda(\br) \sum_{\bp}
\int d\tau \int d\br_1 \langle {\cal T} \hat{\Phi}^\dagger_\alpha(\br,t)
\hat{\psi}_\alpha(\br,t) \hat{\psi}^\dagger_\bp(\br_1,\tau) \rangle
 G^{-1}_{\bp\beta}(\br_1 \tau, \brp t^\prime)\nonumber\\
\Pi_{\alpha\beta}(\br t,\brp t^\prime) &=& \frac{i}{\hbar} \kappa_\alpha \lambda(\br) \sum_{\bp}
\int d\tau \int d\br_1
\langle {\cal T} \hat{\psi}^\dagger_\alpha(\br,t) \hat{\psi}_\alpha(\br,t)
\hat{\Phi}^\dagger_\bp(\br_1,\tau)
\rangle D^{-1}_{\bp\beta}(\br_1 \tau, \brp t^\prime)
\end{eqnarray}

Eqs. (\ref{zero-green}), (\ref{dyson}) and (\ref{self-energy})  are exact and constitute the
non-equilibrium superoperator Green function (NESGF) formalism.

In order to evaluate the self energies perturbatively.
we rewrite the Green functions, Eq. (\ref{green}), in the interaction picture,
\begin{eqnarray}
\label{green-int-pic}
G_{\alpha\beta}(\br t,\brp t^\prime) &=& -\frac{i}{\hbar}\langle {\cal T}
\tilde{\psi}_\alpha(\br,t) \tilde{\psi}^\dagger_\beta(\brp,t^\prime)
{\cal G}_I(t,-\infty) \rangle_0
 \nonumber\\
D_{\alpha\beta}(\br t,\brp t^\prime) &=& -\frac{i}{\hbar}\langle {\cal T}
\tilde{\Phi}(\br,t) \tilde{\Phi}^\dagger(\brp,t^\prime) {\cal G}_I(t,-\infty) \rangle_0
\end{eqnarray}
where ${\cal G}_I(t,-\infty)$ is given by Eq. (\ref{time-int}) with $t_0=-\infty$.
Using Eqs. (\ref{inter-G}),(\ref{time-L}) and (\ref{rhot-1}), the self energies
(\ref{self-energy}) can be also expressed in the interaction picture as,
\begin{eqnarray}
\label{self-energy-int}
\Sigma_{\alpha\beta}(\br t,\brp t^\prime) &=&\! - \frac{i}{\hbar}\kappa_\alpha \lambda(\br) \sum_{\bp}
\int\!\int d\tau d\br_1 \langle {\cal T} \tilde{\Phi}^\dagger_\alpha(\br,t)
\tilde{\psi}_\alpha(\br,t) \tilde{\psi}^\dagger_\bp(\br_1,\tau) {\cal G}_I(t,-\infty) \rangle_0
 G^{-1}_{\bp\beta}(\br_1 \tau, \brp t^\prime)\nonumber\\
\Pi_{\alpha\beta}(\br t,\brp t^\prime) &=&\!\frac{i}{\hbar} \kappa_\alpha\lambda(\br)\!\sum_{\bp}
\int\!\int d\tau d\br_1
\langle {\cal T} \tilde{\psi}^\dagger_\alpha(\br,t) \tilde{\psi}_\alpha(\br,t)
\tilde{\Phi}^\dagger_\bp(\br_1,\tau) {\cal G}_I(t,-\infty)\rangle_0
D^{-1}_{\bp\beta}(\br_1 \tau, \brp t^\prime).
\end{eqnarray}
Equations (\ref{self-energy-int}) together with (\ref{green-int-pic}) constitute  closed form
equations for the self-energies where all the averages are given in the interaction picture,
$<...>_0$, with respect to the non-interacting density matrix. By expanding ${\cal G}_I$
(Eq. \ref{time-int}) perterbatively in ${\tilde{\cal H}_-}^\prime$ we can obtain perturbative
expansion for the self-energies. Each term in the in the expansion can be calculated using Wick's
theorem for superoperators \cite{shaul-pre68} which is given in the Appendix E. This results in
a perturbative series in terms of the zeroth order Green functions.

\section*{III. The Calculation of Molecular Currents}
We have applied the NESGF to study the charge conductivity of a molecular wire attached to two perfectly
conducting leads. In the simplest approach the leads '$a$' and '$b$' are treated as two free electron
reservoirs. Nuclear motions in the molecular region are described as harmonic phonons
which interact with the surrounding elctronic structure and the environment (secondary phonons)
\cite{a-b-nitzan}. We first recast the general Hamiltonian, Eq. (\ref{hamilton}), in a single
electron local basis and partition it as,
\begin{equation}
\label{hamilton1}
H = H_f + H_{int}
\end{equation}
where $H_f$ represents the free, non-interacting electrons and phonons and with no coupling between
molecule and leads,
\begin{equation}
\label{hf}
H_f= \sum_{i,j} E_{i,j} \psi^\dagger_i \psi_j + \sum_{k\in
a,b}\epsilon_k \psi^\dagger_k
\psi_k + \sum_l \Omega_l \phi^\dagger_l \phi_l
+\sum_m \omega_m \phi^\dagger_m \phi_m.
\end{equation}
The indices $(i,j)$ represent the electronic basis states corresponding to the molecule,  $k$
labels the electronic states
in the leads ($a$ and $b$), $l$ denotes primary phonons which intercts with the electrons and
 $m$ denotes the secondary phonons which are coupled to the primary phonons and constitute a thermal bath.
The applied external voltage $V$ maintains a chemical potential diffrence, $\mu_a$-$\mu_b$ = $eV$, between
the two leads and also modifies the single electron energies. In addition it provides an extra
term $\sum_i V_i \psi^\dagger_i \psi_i$ which is included in the zeroth order hamiltonian, $H_f$, by
modifying the single electron energies. The interaction Hamiltonian is given by,
\begin{equation}
\label{hint}
H_{int}= \sum_{k\in a,b;i} \left( V_{ki} \psi^\dagger_k \psi_i
+ h.c. \right)
+\sum_{l,i} \lambda_{li} \Phi_l \psi^\dagger_i \psi_i + \sum_{l,m}
U_{lm} \Phi_l \Phi_m.
\end{equation}
The three terms represent the molecule/lead interaction, coupling of primary phonons with the molecule and the interaction of primary and secndary phonons, respectively.

The total current passing through the  junction can be
expressed in terms of the electron Green functions and the corresponding self
energies. At steady state it is given by (see Appendix B, Eq. \ref{current6}),
\begin{equation}
\label{current}
I_T = \frac{2e}{\hbar}\sum_{ij^\prime} \int \frac{d\omega}{2\pi} \left[
\Sigma_{LR}^{ij^\prime}(\omega)
G_{RL}^{j^\prime i}(\omega) - \Sigma_{RL}^{ij^\prime}(\omega) G_{LR}^{j^\prime i}(\omega) \right].
\end{equation}
The electron Green functions $G^0_{LR}$ and $G^0_{RL}$ correspond to the free Hamiltonian, $H_f$,
and the self-energies $\Sigma_{LR}$ and $\Sigma_{RL}$ represent the effects of all
interactions (Eq. \ref{hint}).

$\Sigma_{\alpha\beta}^{ij}$ has contributions coming from the electron-lead ($\sigma$) and electron-phonon
($\Xi$) interactions,
\begin{equation}
\label{final-self-elec}
\Sigma^{ij}_{\alpha\beta}(\omega) = \sigma^{ij}_{\alpha\beta}(\omega) + \Xi^{ij}_{\alpha\beta}(\omega)
\end{equation}
These are given in Eqs. (\ref{freq-self2}) and (\ref{freq-energy}).
The self energy expressions (\ref{freq-energy}) and (\ref{freq-energy-phonon}) are calculated
perturbatively to second order in the electron-phonon coupling in terms of the zeroth order
Green functions (Eq. \ref{free-gree-basis}).
The simplest expression for current is obtained by substituting Eqs. (\ref{free-gree-basis}), (\ref{freq-self2}) and  (\ref{freq-energy}) in Eq. (\ref{current}).
This zeroth order result can be improved by using the renormalized Green functions
obtained  from the self-consistent solution
of Dyson equation (\ref{dyson}).

In order to solve self-consistently for the
electron Green functions that appear in the current formula, Eq.
(\ref{current}), the self energy is calculated under the Born approximation by replacing the zeroth order Green functions, $G^0_{\alpha\beta}$ and $D^0_{\alpha\beta}$ with the corresponding renormalized Green functions, $G_{\alpha\beta}$ and $D_{\alpha\beta}$, as is commonly done in mode-coupling theories \cite{bouchand,oppenheim}.
This approximation sums an infinite set of non-crossing
diagrams \cite{bickers,meir} that appear in the perturbation expansion of the many body Green function,
$G_{\alpha\beta}$.

Since the electron self-energy (Eq. \ref{freq-energy}) also depends on the phonon Green function,
the phonon self-energy, $\Pi^{ll^\prime}_{\alpha\beta}$, is also required for a self-consistent
solution of the electron Green functions. The phonon self-energy calculated in Appendix C is given by,
\begin{equation}
\label{final-self-phonon}
\Pi_{\alpha\beta}^{ij}(\omega) = \gamma_{\alpha\beta}^{ij}(\omega) +
\Lambda_{\alpha\beta}^{ij}(\omega)
\end{equation}
where $\gamma_{\alpha\beta}^{ij}(\omega)$ (Eq. \ref{freq-self3}) and $\Lambda_{\alpha\beta}^{ij}(\omega)$ (Eq. \ref{freq-energy-phonon})
represent the contributions from the phonon-phonon and the electron-phonon interactions, respectively.

Computing the renormalized electron and phonon Green functions and the corresponding self-energies involves the  self-consistent solution of the following coupled
equations for the Green functions. 
\begin{eqnarray}
\label{coupled}
G_{LR}(\omega) &=&  G^0_{LL}(\omega) \Sigma_{LL}(\omega) G_{LR}(\omega) +
G^0_{LL}(\omega)\Sigma_{LR}(\omega) G_{RR}(\omega) \nonumber\\
G_{RL}(\omega) &=& G^0_{RR}(\omega) \Sigma_{RL}(\omega) G_{LL}(\omega) +
G^0_{RR}(\omega)\Sigma_{RR}(\omega) G_{RL}(\omega) \nonumber\\
G_{LL}(\omega) &=& G^0_{LL}(\omega) + G^0_{LL}(\omega) \Sigma_{LL}(\omega) G_{LL}(\omega) +
G^0_{LL}(\omega) \Sigma_{LR}(\omega) G_{RL}(\omega) \nonumber\\
G_{RR}(\omega) &=& G^0_{RR}(\omega) + G^0_{RR}(\omega)\Sigma_{RL}(\omega) G_{LR}(\omega) +
G^0_{RR}(\omega)\Sigma_{RR}(\omega) G_{RR}(\omega).
\end{eqnarray}
Similarly the equations for the phonon Green functions are obtained by replacing
$G_{\alpha\beta}$ with $D_{\alpha\beta}$ and $\Sigma_{\alpha\beta}$ with $\Pi_{\alpha\beta}$.
Here Green functions corresponding to the free Hamiltonian $G^{0ij}_{\alpha\beta}$ and $D^{0ll^\prime}_{\alpha\beta}$ are given by,
\begin{eqnarray}
\label{free-gree-basis}
G^{0ij}_{\alpha\beta}(\omega) &=&  \frac{\delta_{\alpha\beta}}{\omega \delta_{ij} -\kappa_\alpha E_{ij} +i\eta} \nonumber\\
D^{0ll^\prime}_{\alpha\beta}(\omega) &=&  -\frac{\delta_{\alpha\beta}\delta_{ll^\prime}}
{\omega-\kappa_\alpha\Omega_l + i\eta}
\end{eqnarray}
where we set $\hbar$= 1 and $\eta \to 0$. $E_{ij}$=$E_i$-$E_j$ is the energy difference between
single electron $i$th and $j$th states. $\Omega_l$ denotes the molecular phonon eigenstates.

Once the Green functions $G_{LR}$, $G_{RL}$ and the corresponding self-energies $\Sigma_{LR}$,
$\Sigma_{RL}$ are obtained from the self-consistent solution of Eqs. (\ref{coupled}) together with (\ref{final-self-elec}), formula (\ref{current}) can be used to calculate the total current through the molecular junction.

\section*{Discussion}

In this paper we have developed the NESGF formalism and applied it to the computation
 of molecular current.
The Liouville space time ordering operator provides an elegant way for performing
 calculations in real time, thus avoiding the artificial backward and forward time
 evolution required in Hilbert space (Keldysh loop). Wick's theorem for superoperator is used to compute the self-enrgies  perturbatively to the second order in phonon-electron coupling.
 Equations (\ref{coupled}) have been derived earlier by many authors \cite{caroli1,meir,datta2002}. Recently
Galperin $et$ $al$ \cite{a-b-nitzan} have used a fully self-consistent solution  to
study the influence of different interactions on molecular conductivity for a strong electron-phonon
coupling. The main aim of thepresent work is to demonstrate that by doing calculations in Liouville space one can avoid thebackward/forward time evolution (Keldysh loop) required in Hilbert space. This originates from the fact that in Liouville space both ket and bra evolve forward in time. Thus one can couple the system with two independent fields, "left" and the "right".
This property of Liouville space can be used to construct real (physical) time generating functionals
for the  non-perturbative calculation of the self-energies.

The present model \cite{anitzan3,tikhodeev2001,tikhodeev} ignores electron-electron interactions.
These may be treated using the GW technique \cite{hedin,onida,luie} formulated in terms of the superoperators and extended to non-equilibrium situations. All non-equilibirium observables can be obtained from a single generating functional in terms of "left" and "right" operators.
The retarded (advance) Green function that describe the forward (backward) motion of the
system particle  can also be calculated in terms of the basic Green
functions, $G_{\alpha\beta}$ (see Appendix D).

The NESGF formulation can be also recast in terms of the $+$ and $-$ (rather than $L/R$)
superoperators which are more directly related to observables. This is done in Appendix D. We focused on the primary quantities that are represented in terms of the "left" and "right" superoperators and all  other quantities are obtained as the linear combination of these basic operators.

\section*{Acknowledgment}
The support of the National Science Foundation (Grant No. CHE-0132571 ) and
NIRT (Grant No. EEC 0303389) is gratefully acknowledged. We wish to thank Prof. Wilson Ho for
usefull discussions and Prof. Abraham Nitzan for sending us the preprint of his paper (Ref. 23).

\section*{Appendix A: Superoperator Expressions for the Keldysh Green
Functions}

The standard NEGF theory formulated in terms of the four
Hilbert space Green functions: time ordered $(G^T)$, anti-time ordered $(G^{\tilde
T})$, greater ($G^>$) and lesser $(G^<)$\cite{keldysh,haug-jauho}. These are defined in the
Heisenberg picture as,
\begin{eqnarray}
\label{hilbert-green}
G^T(\bx,\bxp)&\equiv&-\frac{i}{\hbar}\langle T \hat{\psi}(\bx)
\hat{\psi}^\dagger(\bxp) \rangle \nonumber\\
&=& -\frac{i}{\hbar}\theta(t-t^\prime)\langle \hat{\psi}(\bx)
\hat{\psi}^\dagger(\bxp) \rangle
+ \theta(t^\prime-t) \langle \hat{\psi}^\dagger(\bxp) \hat{\psi}(\bx)  \rangle
\nonumber\\
G^{\tilde T}(\bx,\bxp)&\equiv&-\frac{i}{\hbar}\langle \tilde{T}
\hat{\psi}(\bx) \hat{\psi}^\dagger(\bxp) \rangle \nonumber\\
&=& -\frac{i}{\hbar}\theta(t^\prime -t)\langle \psi(\bx)
\hat{\psi}^\dagger(\bxp) \rangle
+ \theta(t- t^\prime) \langle  \hat{\psi}^\dagger(\bxp) \hat{\psi}(\bx) \rangle
\nonumber\\
G^>(\bx,\bxp) &\equiv&-\frac{i}{\hbar}\langle \hat{\psi}(\bx)
\hat{\psi}^\dagger(\bxp) \rangle \nonumber\\
G^<(\bx,\bxp) &\equiv&\frac{i}{\hbar}\langle  \hat{\psi}^\dagger(\bxp)
\hat{\psi}(\bx) \rangle
\end{eqnarray}
These are known as
  $T$ ($\tilde T$) is the Hilbert space time (anti-time)
ordering operator: When applied to a product of
operators, it reorders them in ascending (descending) ) times from
right to left.

The four Green functions that show up naturally in Liouville space are
defined as,
\begin{eqnarray}
\label{liouville-green}
G_{LL}(\bx,\bxp)&=& - \frac{i}{\hbar}\langle{\cal T} \hat{\psi}_L(\bx)
\hat{\psi}^\dagger_L(\bxp)\rangle \nonumber\\
G_{RR}(\bx,\bxp)&=& - \frac{i}{\hbar}\langle{\cal T} \hat{\psi}_R(\bx)
\hat{\psi}^\dagger_R(\bxp)\rangle
\nonumber\\
G_{LR}(\bx,\bxp)&=& - \frac{i}{\hbar}\langle{\cal T} \hat{\psi}_L(\bx)
\hat{\psi}^\dagger_R(\bxp)\rangle
\nonumber\\
G_{RL}(\bx,\bxp)&=& - \frac{i}{\hbar}\langle{\cal T} \hat{\psi}_R(\bx)
\hat{\psi}^\dagger_L(\bxp)\rangle
\end{eqnarray}
${\cal T}$ is the Liouville space time ordering operator, which
rearanges all superoperators
in increasing order of time from right to left.

To establish connection between Liouville space and Hilbert space Green functions
we shall convert superoperators back to ordinary operators \cite{oleg}.
For $G_{LR}$ and $G_{RL}$, we obtain,
\begin{eqnarray}
\label{glr&grl}
G_{LR}(\bx,\bxp) &\equiv& - \frac{i}{\hbar} \mbox{Tr}\{{\cal T}
\hat{\psi}_L(\bx) \hat{\psi}^\dagger_R(\bxp)
\rho_{eq} \}\nonumber\\
&=& - \frac{i}{\hbar} \mbox{Tr}\{\hat{\psi}(\bx) \rho_{eq}
\hat{\psi}^\dagger(\bxp)\}\nonumber\\
&=& - \frac{i}{\hbar} \langle  \hat{\psi}^\dagger(\bxp) \hat{\psi}(\bx) \rangle =
G^<(\bx,\bxp)\nonumber\\
G_{RL}(\bx,\bxp) &\equiv& - \frac{i}{\hbar} \mbox{Tr}\{{\cal T}
\hat{\psi}_R(\bx) \hat{\psi}^\dagger_L(\bxp)
\rho_{eq}\}\nonumber\\
&=& - \frac{i}{\hbar} \mbox{Tr}\{ \hat{\psi}^\dagger(\bxp) \rho_{eq}
\hat{\psi}(\bx) \}\nonumber\\
&=& - \frac{i}{\hbar} \langle \hat{\psi}(\bx) \hat{\psi}^\dagger(\bxp) \rangle =
G^>(\bx,\bxp)
\end{eqnarray}
where $\rho_{eq}$ is the fully interacting many body equilibrium density matrix.

 For $G_{LL}$ and $G_{RR}$ we have two cases,

(i). For $t > t^\prime$, we get,
\begin{eqnarray}
\label{gll&gr1}
G_{LL}(\bx,\bxp) &\equiv& - \frac{i}{\hbar} \mbox{Tr}\{{\cal T}
\hat{\psi}_L(\bx) \hat{\psi}^\dagger_L(\bxp)
\rho_{eq}\}\nonumber\\
&=& - \frac{i}{\hbar} \mbox{Tr}\{\hat{\psi}(\bx) \hat{\psi}^\dagger(\bxp)
\rho_{eq} \}= - \frac{i}{\hbar} \langle \hat{\psi}(\bx) \hat{\psi}^\dagger(\bxp) \rangle
\nonumber\\
G_{RR}(\bx,\bxp) &\equiv& - \frac{i}{\hbar} \mbox{Tr}\{{\cal T}
\hat{\psi}_R(\bx) \hat{\psi}^\dagger_R(\bxp)
\rho_{eq}\}\nonumber\\
&=& - \frac{i}{\hbar} \mbox{Tr}\{\rho_{eq} \hat{\psi}^\dagger(\bxp)
\hat{\psi}(\bx)\}= - \frac{i}{\hbar} \langle  \hat{\psi}^\dagger(\bxp) \hat{\psi}(\bx) \rangle
\end{eqnarray}
 (ii) For the reverse case,  $t < t^\prime $, we get,

\begin{eqnarray}
\label{gll&gr2}
G_{LL}(\bx,\bxp) &\equiv& - \frac{i}{\hbar} \mbox{Tr}\{{\cal T}
\hat{\psi}_L(\bx) \hat{\psi}^\dagger_L(\bxp)
\rho_{eq}\}\nonumber\\
&=& - \frac{i}{\hbar} \mbox{Tr}\{ \hat{\psi}^\dagger(\bxp) \hat{\psi}(\bx)
\rho_{eq} \}=
- \frac{i}{\hbar} \langle  \hat{\psi}^\dagger(\bxp) \hat{\psi}(\bx) \rangle
\nonumber\\
G_{RR}(\bx,\bxp) &\equiv& - \frac{i}{\hbar} \mbox{Tr}\{{\cal T}
\hat{\psi}_R(\bx) \hat{\psi}^\dagger_R(\bxp)
\rho_{eq}\}\nonumber\\
&=& - \frac{i}{\hbar} \mbox{Tr}\{\rho_{eq} \hat{\psi}(\bx)
\hat{\psi}^\dagger(\bxp)\}=
- \frac{i}{\hbar} \langle \hat{\psi}(\bx) \hat{\psi}^\dagger(\bxp)\rangle
\end{eqnarray}
Combining Eqs. (\ref{gll&gr1}) and (\ref{gll&gr2}) we can write,
\begin{eqnarray}
\label{gll&grr}
G_{LL}(\bx,\bxp) &\equiv& - \frac{i}{\hbar} \mbox{Tr}\{{\cal T}
\hat{\psi}_L(\bx) \hat{\psi}^\dagger_L(\bxp)
\rho_{eq}\} \nonumber\\
&=& -\frac{i}{\hbar} \left[ \theta(t-t^\prime) \langle \hat{\psi}(\bx)
\hat{\psi}^\dagger(\bxp) \rangle
- \theta(t^\prime-t) \langle  \hat{\psi}^\dagger(\bxp) \hat{\psi}(\bx) \rangle
\right] \nonumber\\
&=& G^{T}(\bx,\bxp) \nonumber\\
G_{RR}(\bx,\bxp) &\equiv& - \frac{i}{\hbar} \mbox{Tr}\{{\cal T}
\hat{\psi}_R(\bx) \hat{\psi}^\dagger_R(\bxp)
\rho_{eq}\} \nonumber\\
&=& -\frac{i}{\hbar} \left[ \theta(t-t^\prime) \langle
\hat{\psi}^\dagger(\bxp) \hat{\psi}(\bx) \rangle
- \theta(t^\prime-t) \langle  \hat{\psi}(\bx) \hat{\psi}^\dagger(\bxp)  \rangle
\right] \nonumber\\
&=& G^{\tilde T}(\bx,\bxp)
\end{eqnarray}
Eqs. (\ref{glr&grl}) and (\ref{gll&grr}) establish the equivalence of
Hilbert and Liouville space Green functions and they can be summerized as,
\begin{eqnarray}
\label{summery}
G_{LL}(\bx,\bxp) &=& G^T(\bx,\bxp), ~~~~ G_{RR}(\bx,\bxp) = G^{\tilde{T}}(\bx,\bxp) \nonumber\\
G_{LR}(\bx,\bxp) &=& G^<(\bx,\bxp), ~~~~ G_{RL}(\bx,\bxp) = G^>(\bx,\bxp)
\end{eqnarray}

\section*{Appendix B: Superoperator Green Function Expression for the Current }

In this Appendix we present a formal microscopic derivation for the
current flowing through a conductor.
The conductor could be a molecule or a metal or any conducting material
attached to two electrodes held at two different potentials.

   In Hilbert space the charge current-density is given by,
\begin{equation}
\label{current1}
{\bf j}(\br,t)= -\frac{ie\hbar}{2m} \langle \left[ \hat{\psi}^\dagger(\br,t)
{\bf \nabla} \hat{\psi}(\br,t) -
({\bf \nabla} \hat{\psi}^\dagger(\br,t)) \hat{\psi}(\br,t) \right] \rangle
\end{equation}
where $e$ and $m$ are the electron charge and mass, respectively. Eq.
(\ref{current1}) can be also
expressed in a slightly modified form as,
\begin{equation}
\label{current2}
{\bf j}(\br,t)= \frac{ie\hbar}{2m} \left[ \langle  ({\bf \nabla} - {\bf
\nabla}^\prime) \hat{\psi}^\dagger(\br,t)  \hat{\psi}(\brp,t^\prime) \rangle
\right]_{\bxp=\bx}
\end{equation}
where ${\bf \nabla}^\prime$ represents the derivative with respect to
$\brp$.

Using relations (\ref{summery}) the current density can be expressed in
terms of the superoperator Green function as,
\begin{equation}
{\bf j}(\br,t)= -\frac{e\hbar^2}{2m}\left[ ({\bf \nabla} - {\bf
\nabla}^\prime)
G_{LR}(\br t,\brp t^\prime) \right]_{\bxp=\bx}
\end{equation}
At steady state, the Green functions only depend on the time difference $(t-t^\prime)$ and the
total current density $({\bf J}_T)$ becomes time independent.
Transforming to the frequency (energy) domain, the current density per unit energy is,
\begin{equation}
\label{current3}
{\bf j}(\br,E)= -\frac{e\hbar}{2m}\left[ ({\bf \nabla} - {\bf \nabla}^\prime)
G_{LR}(\br \brp, E ) \right]_{\brp=\br},
\end{equation}
and the total current density
\begin{equation}
\label{current30}
{\bf J}_T(\br)=  \int \frac{dE}{2\pi} {\bf j}(\br,E).
\end{equation}
Eq. (\ref{current30}) provides a recipe for calculating the current profile
across the conductor once the
Green function $G_{LR}$ is known from  the self-consistent solution of
the Dyson equation. For computing the total current passing through the conductor, Eq.
(\ref{current3}) can be expressed
in the form of Eq. (\ref{current}). In order to get the total current
per unit energy $(I_T(E))$  passing between electrode/conductor we need to integrate the current density
over the surface area of the conductor-electrode contact.
\begin{equation}
\label{total-current}
I_T(E) = \int_s {\bf j}(\br,E) \cdot {\bf \hat{n}} dS =
\int {\bf \nabla} \cdot  {\bf j}(\br,E)  d\br,
\end{equation}
where ${\bf \hat{n}}$ is the unit vector normal to surface $S$.
Substituting into Eq. (\ref{total-current}) from (\ref{current3}), we
get
\begin{equation}
\label{current4}
I_T(E)= -\frac{e\hbar}{2m} \mbox{Tr} \left[ (\nabla^2 -
{\nabla^\prime}^2) G_{LR}(\br \brp, E ) \right].
\end{equation}
In general, a conductor-electrode system can be described by the
Hamiltonain
\begin{equation}
H = H_0 + H_{int}
\end{equation}
where $H_0$ represents the non-interacting part,
\begin{equation}
H_0 = \int d\br \psi^\dagger(\br) h_0(\br) \psi(\br)
\end{equation}
where $h_0(\br)$= $-\frac{\hbar^2}{2m} \nabla^2$
and all the interaction terms (conductor-electrode, electron-phonon)are
included in $H_{int}$.
The total current per unit energy, Eq. (\ref{current4}), is
\begin{equation}
\label{current5}
I_T(E)= -\frac{e}{\hbar}\mbox{Tr} \left[ (h_0(\br) - h_0^*(\brp))
G_{LR} (\br \brp,E) \right].
\end{equation}
The Dyson equations for the retarded Green function (see Appendix D, Eq. \ref{ndyson}),
in frequency (energy) can be expressed in the matrix form as,
\begin{equation}
\label{nndyson-r}
h_0 G_r = E G_r - \mbox I - \sigma_r G_r
\end{equation}
where $\mbox{I}$ is the identity matrix and $\Sigma_r$ is the retarded self-energy, Eq. (\ref{ret-adv-energy}). $E$=$\hbar\omega$ is a number. Henceforth we write all the expressins
in the matrix notation.
Taking the complex conjugate of (\ref{nndyson-r}), we obtain Dyson equation for the advanced Green
function,
\begin{equation}
\label{nndyson-a}
G_a  h_0= E G_a - \mbox I - G_a \sigma_a
\end{equation}
with the corresponding advanced self-energy, $\Sigma_a$.
From the matrix Dyson equation (\ref{ndyson}), we also have the relation,
\begin{equation}
\label{check}
G_{LR} = G_r \Sigma_{LR} G_a
\end{equation}
Using the relations (\ref{nndyson-r})-(\ref{check}), it is easy to see that
\begin{equation}
\label{relation-1}
h_0 G_{LR} - G_{LR} h_0 = G_{LR} \Sigma_a + G_r \Sigma_{LR} - \Sigma_r
G_{LR} - \Sigma_{LR} G_a
\end{equation}
Substituting this in Eq. (\ref{current5}), the total current per unit
energy becomes,
\begin{equation}
\label{current6}
I_T(E) = \frac{2e}{\hbar} \mbox{Tr} [\Sigma_{LR}(E) G_{RL}(E) -
\Sigma_{RL}(E) G_{LR}(E).
\end{equation}
Where a factor of 2 is introduce to account for the spin degeneracy.

We have calculated the total current
in real space. In practice the Green functions and the self-energy matrices are
calculated in an electronic basis ($i,j$).
The total current through the conductor is obtained by integrating
Eq. (\ref{current6}) over energy resulting in Eq. (\ref{current}).

\section*{Appendix C: Self-energies for Superoperator Green Functions }
The basic quantities required for describing the coupled
molecule-lead system are the one particle
electron and the phonon Green functions.
Following the steps outlined in Sec. II, the time development
for various superoperators ( Heisenberg equations) is (all primed indices should be summed over),
\begin{eqnarray}
\label{super-dyn3}
i\hbar \kappa_\alpha \frac{\partial}{\partial t} \hat{\psi}_{i\alpha}(t)
&=& E_{ij^\prime} \hat{\psi}_{j^\prime\alpha}(t) + V_{k^\prime i}
\hat{\psi}_{k^\prime\alpha}(t)
+ \lambda_{l^\prime i}
\hat{\Phi}_{l^\prime\alpha}(t)\hat{\psi}_{i\alpha}(t)\nonumber\\
-i\hbar\kappa_\alpha\frac{\partial}{\partial t} \hat{\Phi}_{l\alpha}(t) &=&
\lambda_{i^\prime l}
\hat{\psi}^\dagger_{i^\prime\alpha}(t) \hat{\psi}_{i^\prime\alpha}(t) +
\Omega_l \hat{\Phi}_{l\alpha}(t) +
U_{lm^\prime} \hat{\Phi}_{m^\prime\alpha}(t)\nonumber\\
i\hbar\kappa_\alpha\frac{\partial}{\partial t}
\hat{\psi}_{k\alpha}(t) &=&
\epsilon_k \hat{\psi}_{k\alpha}(t) + V_{ki^\prime}
\hat{\psi}_{i^\prime}(t)\nonumber\\
-i\hbar\kappa_\alpha\frac{\partial}{\partial t}
\hat{\Phi}_{m\alpha}(t) &=&
\omega_m \hat{\Phi}_{m\alpha}(t) + U_{l^\prime m} \hat{\Phi}_{l^\prime}(t),
\end{eqnarray}
where $\Phi_{m\alpha}=\phi^\dagger_{m\alpha}+\phi_{m\alpha}$.
Using Eqs. (\ref{super-dyn3}) it is straightforward to write
the matrix Dyson equation (\ref{dyson}) for the electron and phonon Green functions defined as,
\begin{eqnarray}
\label{dyson-basis}
G^{ij}_{\alpha\beta}(\bx,\bxp) &=& -\frac{i}{\hbar} \langle {\cal T} \psi_{i\alpha}(\br,t)
\psi^\dagger_{j\beta}(\brp,t^\prime) \rangle \nonumber\\
D^{ll^\prime}_{\alpha\beta}(\bx,\bxp) &=& -\frac{i}{\hbar} \langle {\cal T} \Phi_{l\alpha}(\br,t)
\Phi^\dagger_{l^\prime\beta}(\brp,t^\prime) \rangle,
\end{eqnarray}
with the corresponding self-energy matrix elements,
\begin{eqnarray}
\label{self-energy2}
\Sigma^{ij}_{\alpha\beta}(t,t^\prime) &=&
-\frac{i}{\hbar}\kappa_\alpha \sum_{\bp,j^\prime}
 \int d\tau \left[ \sum_{l^\prime,}\lambda_{l^\prime i} \langle {\cal
T} \hat{\Phi}_{l^\prime\alpha}(t)
\hat{\psi}_{i\alpha}(t) \hat{\psi}^\dagger_{j^\prime\bp}(\tau) \rangle
\right.\nonumber\\
&+& \left. \sum_{k^\prime} V_{k^\prime i}\langle {\cal T} \hat{\psi}_{k^\prime\alpha}(t)
\hat{\psi}^\dagger_{j^\prime\bp}(\tau)\rangle \right] {G^{j^\prime
j}_{\bp\beta}}^{-1}(\tau, t^\prime)\equiv \Xi^{ij}_{\alpha\beta}(t,t^\prime)+
\sigma^{ij}_{\alpha\beta}(t,t^\prime) \nonumber\\
\Pi^{ll^\prime}_{\alpha\beta}(t,t^\prime) &=& \frac{i}{\hbar} \int
d\tau \sum_{\bp,l^{\prime\prime}}
\left[ \sum_{m^\prime} U_{lm^\prime} \langle {\cal T}
\hat{\Phi}_{m^\prime\alpha}(t)
\hat{\Phi}^\dagger_{l^{\prime\prime}\bp}(\tau) \rangle \right.\nonumber\\
&+& \left. \sum_{i^\prime} \lambda_{li^\prime} \langle {\cal T}
\hat{\psi}^\dagger_{i^\prime\alpha}(t)
\hat{\psi}_{i^\prime\alpha}(t) \hat{\Phi}^\dagger_{l^{\prime\prime}\bp}(\tau)
\rangle \right]  {D^{l^{\prime\prime} l^\prime}_{\bp\beta}}^{-1} (\tau,t^\prime)
\equiv \gamma^{ll^\prime}_{\alpha\beta}(t,t^\prime)+ \Lambda^{ll^\prime}_{\alpha\beta}(t,t^\prime)
\end{eqnarray}
The two terms in the electron self energy represent the contributions from
the phonon-electron ($\Xi$) and molecule - lead ($\sigma$)interactions. Similarly, the
phonon self energy has contributions from the electron-phonon ($\Lambda$) and
the primary-secondary phonon ($\gamma$) couplings. The self energy due to the
molecule-lead coupling can be calculated
exactly. To that end we need to obtain 
the quantity $\langle {\cal T} {\tilde \psi}_{k^\prime\alpha}(t)
\psi^\dagger_{j^\prime\bp}(\tau)\rangle$.
By multiplying the third equation in (\ref{super-dyn3}) by
$\psi^\dagger_{j^\prime\beta^\prime}(\tau)$
from the left and from the right, taking trace and subtracting, we
get (here primed indices are not summed over),
\begin{eqnarray}
\label{mol-lead1}
\left(i\hbar\kappa_\alpha \frac{\partial}{\partial t} -
\epsilon_{k^\prime}\right) \langle {\cal T} \hat{\psi}_{k^\prime\alpha}(t)
\hat{\psi}^\dagger_{j^\prime\beta^\prime}(\tau) \rangle &=&
\sum_{i^\prime} V_{k^\prime i^\prime}\langle {\cal T}
\hat{\psi}_{i^\prime\alpha}(t)
\hat{\psi}^\dagger_{j^\prime\beta^\prime}(\tau) \rangle \nonumber\\
\Rightarrow ~~~~~~~~ \langle {\cal T} \hat{\psi}_{k^\prime\alpha}(t)
\hat{\psi}^\dagger_{j^\prime\beta^\prime}(\tau) \rangle &=&
i\hbar \sum_{i^\prime} V_{k^\prime i^\prime} g_{k^\prime}(t)
G^{i^\prime j^\prime}_{\alpha\bp}(t,\tau)
\end{eqnarray}
where $g_k(t)= \left( i\hbar\kappa_\alpha \frac{\partial}{\partial t}
- \epsilon_{k^\prime} \right)^{-1}$. Substituting expression
(\ref{mol-lead1})
in Eq. (\ref{self-energy2}) gives for the molecule-lead self energy,
\begin{equation}
\label{mol-lead}
\sigma^{ij}_{\alpha\beta}(t,t^\prime) = \kappa_{\alpha}
\delta_{\alpha\beta} \sum_{k^\prime\in a,b} V_{k^\prime i} V_{k^\prime j}
g_{k^\prime}(t) \delta(t-t^\prime)
\end{equation}
Similarly, the contribution to the phonon self energy from the
interaction with secondary phonons
can be calcualted exactly,
\begin{equation}
\label{sec-phonon}
\gamma^{ll^\prime}_{\alpha\beta} = -\kappa_\alpha
\delta_{\alpha\beta}
\sum_{m^\prime} U_{lm^\prime} U_{l^\prime m^\prime}
g^\prime_{m^\prime}(t)
\delta(t-t^\prime)
\end{equation}
where $g^\prime_{m^\prime}(t)= \left( i\hbar\kappa_\alpha
\frac{\partial}{\partial t} + \omega_{m^\prime} \right)^{-1}$.
At steady state all Green functions and self-energies depend only on the time difference
$(t_1-t_2)$ and it is very convenient to express them in the frequency space.
The self energy contributions due to molecule-lead
$(\sigma^{ij}_{\alpha\beta})$ and
phonon-phonon ($\gamma^{l^\prime}_{\alpha\beta}$) interactions, Eqs.
(\ref{mol-lead}) and
(\ref{sec-phonon}), can be represented in frequency space as,
\begin{equation}
\label{freq-self2}
\sigma^{ij}_{\alpha\beta}(\omega) = \kappa_{\alpha}
\delta_{\alpha\beta}
\sum_{k^\prime}\frac{V_{k^\prime i} V_{k^\prime j}}{\kappa_\alpha
\omega-\epsilon_{k^\prime} +i\eta}
\end{equation}
\begin{equation}
\label{freq-self3}
\gamma^{ll^\prime}_{\alpha\beta}(\omega) = - \kappa_{\alpha}
\delta_{\alpha\beta}
\sum_{m^\prime} \frac{U_{lm^\prime} U_{l^\prime
m^\prime}}{\kappa_\alpha \omega + \omega_{m^\prime} +i\eta}
\end{equation}
where $\eta \to 0$. However in real calculations it is a common
practice to calculate self energies $\sigma_{\alpha\beta}$ and $\gamma_{\alpha\beta}$ in the wide band
approximation implying that the real parts of the self energies can be ignored and the imaginary parts
are considered as frequency independent. Eqs. (\ref{freq-self2}) and (\ref{freq-self3}) then reduce to simpler forms,
\begin{eqnarray}
\sigma^{ij}_{\alpha\beta} &=& \kappa_{\alpha} \delta_{\alpha\beta}
\frac{i}{2}\Gamma^{ij} \nonumber\\
\gamma^{ll^\prime}_{\alpha\beta} &=& - \kappa_{\alpha}
\delta_{\alpha\beta}
\frac{i}{2}\tilde{\Gamma}^{ll^\prime}
\end{eqnarray}
where $\Gamma^{ij}= 2\pi \sum_{k^\prime} V_{k^\prime i} V_{k^\prime j}$
and
$\tilde{\Gamma}^{ll^\prime} = 2\pi \sum_{m^\prime} U_{m^\prime l} V_{m^\prime
l^\prime}$.

  The phonon contribution to the electronic self energy is
obtained perturbatively in the phonon-electron coupling.
We recast the phonon contribution (first term on the RHS of Eq.
(\ref{self-energy2}) for $\Sigma^{ij}_{\alpha\beta}$) in the interaction
picture by writing,
\begin{equation}
\label{inter-self-1}
\langle {\cal T} \hat{\Phi}_{l\alpha}(t) \hat{\psi}_{i\alpha}(t)
\hat{\psi}^\dagger_{j\beta}(t^\prime) \rangle =
\langle {\cal T} \tilde{\Phi}_{l\alpha}(t)\tilde{\psi}_{i\alpha}(t)
\tilde{\psi}^\dagger_{j\beta}(t^\prime) {\cal G}_I(t,-\infty)\rangle_0
\end{equation}
where
\begin{eqnarray}
\label{new-time-int}
{\cal G}_I(t,-\infty)&=& \mbox{exp}\left\{-\frac{i}{\hbar} \int
d\tau \sum_{i^\prime\ap}\kappa_\ap\left[
\sum_{l^\prime}\lambda_{l^\prime i^\prime}
\tilde{\Phi}_{l^\prime \ap}(\tau)
\tilde{\psi}^\dagger_{i^\prime\ap}(\tau) \tilde{\psi}_{i^\prime\ap}(\tau)
\right.\right. \nonumber\\
&+& \left.\left. \sum_{k^\prime} V_{k^\prime i^\prime}
(\tilde{\psi}^\dagger_{k^\prime \ap}(\tau)
\tilde{\psi}_{i^\prime\ap}(\tau) + \tilde{ \psi}^\dagger_{i^\prime
\ap}(\tau) \tilde{\psi}_{k^\prime \ap}(\tau)) \right]\right\}.
\end{eqnarray}
Substituting (\ref{new-time-int}) in Eq. (\ref{inter-self-1}),
expanding the exponential
to first order in $\lambda_{li}$ and using Wick's theorem for
superoperators \cite{shaul-pre68} we obtain,
\begin{eqnarray}
\label{43}
\langle {\cal T} \hat{\Phi}_{l\alpha}(t) \hat{\psi}_{i\alpha}(t)
\hat{\psi}^\dagger_{j\beta}(t^\prime) \rangle &=&
-\hbar^2 \sum_{l^\prime i^\prime \ap} \kappa_\ap \lambda_{l^\prime
i^\prime}
\int d\tau D^{0ll^\prime}_{\alpha\ap}(t,\tau)\left[
G^{0ij}_{\alpha\beta}(t,t^\prime)
G^{0i^\prime i^\prime}_{\ap\ap}(\tau,\tau^+) \right.\nonumber\\
&+&  \left. G^{0ii^\prime}_{\alpha\ap}(t,\tau) G^{0i^\prime
j}_{\ap\beta}(\tau,t^\prime) \right].
\end{eqnarray}
Here the superscript $'0'$ represents the trace with respect to the
non-interacting density matrix. The zeroth order Green functions are given
in Eq. (\ref{free-gree-basis}).
The terms coming from the lead-molecule coupling ($V_{ki}$) vanish
because they are odd in creation and anihilation operators.
Substituting (\ref{43}) in Eq. (\ref{self-energy2})gives for the phonon
contribution to the self-energy,
\begin{eqnarray}
\label{self-1}
\Xi^{ij}_{\alpha\beta}(t,t^\prime) &=& i\hbar \sum_{l_1 l_2}
\kappa_\alpha
\lambda_{l_1 i} \left[ \kappa_\beta \lambda_{l_2 j} D^{0l_1
l_2}_{\alpha\beta}(t,t^\prime) G^{0ij}_{\alpha\beta}(t,t^\prime)
\right.\nonumber\\
&+& \left. \delta_{ij} \delta_{\alpha\beta} \delta(t-t^\prime)
\sum_{i_1\ap} \lambda_{l_2 i_1} \kappa_\ap
\int d\tau D^{0l_1l_2}_{\alpha\ap}(t,\tau) G^{0i_1
i_1}_{\ap\ap}(\tau,\tau^+)\right]
\end{eqnarray}
In the derivation of (\ref{self-1}), we have used the identity,
\begin{equation}
\int d\tau \sum_{\ap j^\prime} G^{0ij^\prime}_{\alpha\bp}(t,\tau)
{G^{-1}}^{0j^\prime j}_{\bp\beta}(\tau,t^\prime) = \delta_{\alpha\beta}
\delta_{ij} \delta(t-t^\prime)
\end{equation}
Similarly the contribution of the electron-phonon interaction to the
phonon self energy ( second
term in Eq. (\ref{self-energy2}) for $\Pi^{ij}_{\alpha\beta}$) can be
obtained perturbatively. To the
second order in phonon-electron coupling, we obtain,
\begin{equation}
\Lambda^{ll^\prime}_{\alpha\beta}(t,t^\prime)= -i\hbar\sum_{ij}
\kappa_\alpha \kappa_\beta
\lambda_{li} \lambda_{l^\prime j} \left[
G^{0ji}_{\beta\alpha}(t^\prime,t) G^{0ij}_{\alpha \beta}(t,t^\prime)
+ G^{0ii}_{\alpha\alpha}(t,t^+) G^{0jj}_{\beta
\beta}(t^\prime,{t^\prime}^+) \right]
\end{equation}
To second order in electron-phonon coupling, the electronic self energy
depends on both
the electron and phonon green functions while the phonon self energy
contains only the electron
Green functions.

At steady state we shift to the frequency domain and obtain,
\begin{eqnarray}
\label{freq-energy}
\Xi^{ij}_{\alpha\beta}(\omega) &=& i\hbar \sum_{l_1 l_2}
\kappa_\alpha \kappa_\beta
\lambda_{l_1 i}\lambda_{l_2 j} \int \frac{d\omega^\prime}{2\pi}
 D^{0l_1 l_2}_{\alpha\beta}(\omega^\prime)
~G^{0ij}_{\alpha\beta}(\omega-\omega^\prime) \nonumber\\
&+& \delta_{ij} \delta_{\alpha\beta} \sum_{l_1,l_2,i_1,\ap}
\kappa_\ap \lambda_{l_1 i} \lambda_{l_2 i_1} \rho_{i_1i_1}^0
D^{0l_1l_2}_{\alpha\ap}(\omega=0)
 \end{eqnarray}
where
\begin{equation}
\rho^0_{ii} \equiv i\hbar G^{ii}_{\alpha\alpha}(t=0) = i \int
\frac{dE}{2\pi} G^{ii}_{\alpha\alpha}(E)
\end{equation}
The phonon self-energy becomes,
\begin{eqnarray}
\label{freq-energy-phonon}
\Lambda^{ll^\prime}_{\alpha\beta}(\omega) &=& -i\hbar
\sum_{ij}\kappa_\alpha \kappa_\beta
\lambda_{li}\lambda_{l^\prime j}\int \frac{d\omega^\prime}{2\pi}
G^{0ij}_{\alpha \beta}(\omega^\prime)
G^{0ji}_{\beta\alpha}(\omega^\prime-\omega)\nonumber\\
&+& \frac{i}{\hbar} \sum_{ij}\kappa_\alpha \epsilon_\beta
\lambda_{li} \lambda_{l^\prime j}
 \rho^0_{ii} \rho^0_{jj} \delta(\omega=0)
\end{eqnarray}

\section*{Appendix D: Dyson Equations in the +/- Representation}
 In this Appendix we define the retarded and advance Green's functions and the
corresponding self energies and relate them to the basic Green functions and
self energies obtained in Appendix C.
From definitions (\ref{ret-adv}), the Liouville space retarded ($G_r$)
and advance ($G_a$)
Green functions are defined as,
\begin{equation}
\label{gij-ret}
G^{ij}_r(t,t^\prime) \equiv  -\frac{i}{\hbar}\langle {\cal T} \psi_{i+}(t)
\psi^\dagger_{j-}(t^\prime) \rangle
\end{equation}
\begin{equation}
\label{gij-adv}
G^{ij}_a(t,t^\prime) \equiv  -\frac{i}{\hbar}\langle {\cal T}
\psi_{i-}(t^\prime) \psi^\dagger_{j+}(t)  \rangle.
\end{equation}
We further introduce the correlation function,
\begin{equation}
\label{gij-corr}
G_c^{ij}(t,t^\prime) \equiv -\frac{2i}{\hbar} \langle {\cal T}
\psi_{i+}(t^\prime) \psi^\dagger_{j+}(t)  \rangle
\end{equation}
It follows from Eq. (\ref{ret-adv1}) that there are only three Green functions in the $+/-$
representation. These are given by Eqs. (\ref{gij-ret})- (\ref{gij-corr}).
Using Eq. (\ref{def-2}) these can be represented in terms of the basic
Green functions (\ref{green}) as,
\begin{eqnarray}
\label{green-basic}
G^{ij}_r(t,t^\prime) &=& \frac{1}{2}\left[ G^{ij}_{LL}(t,t^\prime) -
G^{ij}_{LR}(t,t^\prime)
+G^{ij}_{RL}(t,t^\prime) - G^{ij}_{RR}(t,t^\prime)\right]\nonumber\\
&=& G^{ij}_{LL}(t,t^\prime) - G^{ij}_{LR}(t,t^\prime) \nonumber\\
G^{ij}_a(t,t^\prime) &=& \frac{1}{2}\left[ G^{ij}_{LL}(t,t^\prime) -
G^{ij}_{RR}(t,t^\prime)
-G^{ij}_{RL}(t,t^\prime) + G^{ij}_{LR}(t,t^\prime)\right]\nonumber\\
&=& - G^{ij}_{RR}(t,t^\prime) + G^{ij}_{LR}(t,t^\prime) =
G^{ij}_{LL}(t,t^\prime)
- G^{ij}_{RL}(t,t^\prime)\nonumber\\
G^{ij}_c (t,t^\prime) &=& \frac{1}{2} \left[ G_{LL}^{ij}(t,t^\prime) +
G_{RR}^{ij}(t,t^\prime)
+ G_{LR}^{ij}(t,t^\prime) + G_{RL}^{ij}(t,t^\prime)\right]\nonumber\\
&=&  G_{LL}^{ij}(t,t^\prime) + G_{RR}^{ij}(t,t^\prime)
\end{eqnarray}
where we have used the identity $G_{LL}+G_{RR}=G_{LR}+G_{RL}$ which can
be varified using Eq. (\ref{ret-adv1}).
A Dyson equation corresponding to $G_r$, $G_a$ and $G_c$ can be
obtained from Eqs. (\ref{dyson})  using unitary transformation,
\begin{equation}
G = S \bar{G} S^{-1}
\end{equation}
where $G$ represents the matrix
\begin{equation}
G = \left(
\begin{array}{cc}
0 & G_a\\
G_r & G_c
\end{array}
\right)
\end{equation}
and
\begin{equation}
S = \frac{1}{\sqrt{2}}\left(
\begin{array}{cc}
1 & -1 \\
1 & 1
\end{array}
\right)
\end{equation}
The transformed Dyson equation (\ref{dyson}) reads
\begin{equation}
\label{ndyson}
G = G^0 + G^0\tilde{\Sigma} G
\end{equation}
and the corresponding self energy matix reduces to
\begin{equation}
\tilde{\Sigma} = \left(
\begin{array}{cc}
\Sigma_c & \Sigma_r\\
\Sigma_a & 0
\end{array}
\right)
\end{equation}
with the matrix elements given by,
\begin{eqnarray}
\label{ret-adv-energy}
\Sigma_r^{ij}(t,t^\prime) &=&  \Sigma_{LL}^{ij}(t,t^\prime) +
\Sigma_{LR}^{ij}(t,t^\prime)\nonumber\\
\Sigma_a^{ij}(t,t^\prime) &=&  \Sigma_{RR}^{ij}(t,t^\prime) +
\Sigma_{LR}^{ij}(t,t^\prime) \nonumber\\
\Sigma_c^{ij}(t,t^\prime) &=&  \Sigma_{RR}^{ij}(t,t^\prime) +
\Sigma_{LL}^{ij}(t,t^\prime)
\end{eqnarray}
Similar relations also hold for the phonon Greens functions and self
energies.

 Using (\ref{ret-adv-energy}) and (\ref{green-basic}), the
retarded self energies
 for electron and phonon Green functions (retarded) coming from the
electron-phonon coupling
 is obtained as,
\begin{eqnarray}
\Xi^{ij}_r (\omega) &=& i \hbar \sum_{ll^\prime}
\lambda_{l^\prime i} \lambda_{lj}
\int \frac{d\omega^\prime}{2\pi} \left[
D_r^{0ll^\prime} (\omega^\prime) G_r^{0ij}(\omega-\omega^\prime) +
D_r^{0ll^\prime} (\omega^\prime) G_{LR}^{0ij}(\omega-\omega^\prime)
\right. \nonumber\\
&+& \left. D_{LR}^{0ll^\prime} (\omega^\prime)
G_r^{0ij}(\omega-\omega^\prime) \right] \nonumber\\
\Lambda^{ll^\prime}_r(\omega) &=& -i\hbar \sum{ij} \lambda_{li}
\lambda_{l^\prime j}
\int \frac{d\omega^\prime}{2\pi}
\left[G_{LR}^{0ij}(\omega^\prime)G_a^{0ij}(\omega-\omega^\prime)
\right.\nonumber\\
&+&\left. G_r^{0ij}(\omega^\prime)\left(
G_{RL}^{0ij}(\omega-\omega^\prime)
+ G_a(\omega-\omega^\prime) \right)\right]
\end{eqnarray}
Similarly the retarded self energies due to the lead and secondary
phonons can be written in the wide band
limit as
\begin{equation}
\sigma_r^{ij} = \frac{i}{2} \Gamma^{ij}~~~~ \mbox{and}~~~~
\gamma_r^{ll^\prime} = -\frac{i}{2} \tilde{\Gamma}^{ll^\prime}
\end{equation}
where $\Gamma^{ij}$ includes contributions from both the leads $a$ and
$b$, $i$$e$, $\Gamma^{ij}$=
$\Gamma^{ij}_a +$ $\Gamma^{ij}_b$.

\section*{Appendix E: Wick's Theorem for Superoperators}
Wick's theorem for superoperators was formulated in Ref. \cite{shaul-pre68}.
Using Eqs. (\ref{def-2}) and (\ref{commutation0}), it can be shown
that similar to the $L$ and $R$ superoperators, the commutator of
"+" and "-" boson superoperators are also numbers. Thus boson
superoperators follow Gaussian statistics and Wick's theorem holds for both
the $L$, $R$ and "+", "-" representations. However for fermi superoperators life is
more complicated. The anticommutator corresponding to only the "left" or the
"right" fermi superoperators are numbers
but that for the "left" and "right" superoperators, in
general, is not a number. Thus the fermi superoperators are not Gaussian.
However, since the left and right superoperators always
commute, the following Wick's theorem \cite{shaul-pre68} can be applied to
the time ordered product of any number of "left" and "right"
superoperators,$e.$$g.$,
\begin{equation}
\langle {\cal T} A_{i_1\nu_1}(t_1)
A_{i_2\nu_2}(t_2)...A_{i_n\nu_n}(t_n) \rangle_0 =
\sum_{p} \langle {\cal T} A_{i_a\nu_a}(t_a) A_{i_b\nu_b}(t_b) \rangle_0
...
\langle {\cal T} A_{i_p\nu_p}(t_p) A_{i_q\nu_q}(t_q) \rangle_0.
\end{equation}
Here $A_{i_n\nu_n}$, $\nu_n=L,R$, represents either a boson or a
fermion superoperator.
$i_a\nu_a...i_q\nu_q$ is a permutation of $i_1\nu_1...i_n\nu_n$ and sum
on $p$ runs over all possible permutations, keeping the time ordering.
In case of fermions, each term should be multiplied by $(-1)^P$, where
 $P$ is the number of permutations of superoperators required to put
them into a perticular order.
 Only permutaions among either "left" or among "right" superoperators
count in $P$. The permutations
 among "L" and "R" operators leave the product unchanged.

\vspace{.3cm}
\noindent $\dagger$ email: uharbola@uci.edu \\
$*$ email: smukamel@uci.edu

\end{document}